# New ID-based Multi-proxy Multi-Signcryption Scheme from Pairings

Sunder Lal and Tej Singh
Department of Mathematics, IBS, Khandari
Agra-282002 (UP)-INDIA
E-mail:- sunder_lal2@rediffmail.com, tej_singh2ibs@yahoo.co.in

**Abstract:** This paper presents an identity based multi-proxy multi-signcryption scheme from pairings. In this scheme a proxy signcrypter group could authorized as a proxy agent by the cooperation of all members in the original signcrypter group. Then the proxy signcryption can be generated by the cooperation of all the signcrypters in the authorized proxy signcrypter group on the behalf of the original signcrypter group. As compared to the scheme of Liu and Xiao, the proposed scheme provides public verifiability of the signature along with simplified key management.

## 1 Introduction

In 1996, Mambo et al [12] first introduced the concept of a proxy signature scheme, which allows the original signer to delegate his signature power to a designed signer, called the proxy signer. Then the proxy signer is able to sign on the behalf of the original signer. Several other multi proxy signature schemes [1], [2], [3], [4], [5], [6], [15] were proposed. In 2001, Hwang et al [2] first proposed the concept of multi-proxy multi-signature scheme.
In a multi-proxy multi-signature scheme, only the cooperation of all the signers in the authorized proxy group can generate the proxy signature on the behalf of the original signer group.

A new type of cryptographic primitive called "signcryption" which combines a function of digital signature scheme with a symmetric key encryption algorithm was introduced by Zheng in [16]. Signcryption not only provides authenticity and confidentiality in a single step, but also gives more efficient computations than traditional signature-then-encryption. Steinfiel and Zheng [14] and Malone-Lee and Mao [11] proposed efficient signcryption schemes that are based on integer factorization and using RSA.

An identity –based cryptosystem is a novel type of cryptographic scheme proposed by Shamir [13], which enables any pair of users to communicate securely, and to verify each other's signatures without exchanging public or private keys, without keeping any key directories and without using the services of any third party. Problem with the traditional public key cryptosystems (PKCs) are the high cost of the infrastructure needed to manage and authenticate public keys, and the difficulty in managing multiple communities. In an ID-based PKC, everyone's public keys are predetermined by information that uniquely identifies them, such as their email address. There is no need for any public key certificate. A trusted key generation center (KGC) generates the private keys of the entities in the group using their public key. First ID-based signcryption scheme was proposed by Malon-Lee [10] in 2002, Libert and Quisquater [8] proposed a provably secure identity-based signcryption schemes from pairing. Several other ID-based signcryption schemes were proposed. Li et al [7] have proposed a proxy signcryption. However, none of existing signcryption schemes is multi-proxy multi-signcryption scheme. In 2005, Liu Jun Bao and Xiao Guo-Zhen [9] proposed multi-proxy multi-signcryption scheme from pairing.

In this paper we propose an ID-based multi-proxy multi-signcryption scheme from pairing. In this scheme, only the cooperation of all members in the original signcrypter group could authorize a proxy group as his proxy agent. Then only the cooperation of all the signcrypters in the authorized proxy group can generate the proxy signcryption on the behalf of the original signcrypter group. As compared to the scheme in [9] the proposed scheme provides the public verifiability of the signature. Further the scheme is ID-based therefore; the key management problem is simplified.

The rest of the paper is organized as follows. Some definitions and preliminary work are given in section 2, Liu and Xiao scheme is described in section 3, and proposed scheme is given in section 4. The security of the scheme is discussed in section 5. Finally, the conclusions are given in section 6.

## 2    Preliminary Works

In this section, we briefly describe the basic definition and properties of the bilinear pairing.

### 2.1    Bilinear Pairings

Let $G_1$ be a cyclic additive group generated by $P$, whose order is a prime $q$, and $G_2$ be a cyclic multiplicative group of the same order $q$. Let a, b be elements of $Z_q^*$. A bilinear pairings is a map $e: G_1 \times G_1 \to G_2$ with the following properties:

1) Bilinearity: $e(aP, bQ) = e(P,Q)^{ab}$.

2) Non-degeneracy: There exists P and Q such that $e(P,Q) \neq 1$.

3) Computability: There is an efficient algorithm to compute $e(P,Q)$ for all $P, Q \in G_1$

The security of our scheme described here relies on the hardness of the following problems:

**DBDHP:** Given two groups $G_1$ and $G_2$ of the same prime order q, a bilinear map $e : G_1 \times G_1 \rightarrow G_2$ and a generator P of $G_1$ the Decisional Bilinear Diffie-Hellman problem (DBDHP) in ($G_1$, $G_2$, e) is to decide whether $h = e(P,P)^{abc}$ given (P,aP,bP,cP) and an element $h \in G_2$.

**CBDHP:** Given two groups $G_1$ and $G_2$ of the same prime order q, a bilinear map $e : G_1 \times G_1 \rightarrow G_2$ and a generator P of $G_1$, the Computational Bilinear Diffie-Hellman problem (CBDHP) in ($G_1, G_2, e$) is to compute $h = e(P,P)^{abc}$ given (P,aP,bP,cP).

No algorithm is known to be able to solve any of them so far, through **DBDHP** is no harder than **CBDHP**.

## 3  Liu - Xiao Scheme

This scheme consists of four phases:

**System Initialization:** Let $A_1 \ldots\ldots A_n$ be n original signcrypters. Each $A_i$ chooses secret key $x_{ai} \in Z_q^*$ and computes public key $y_{ai} = x_{ai}P$.

Let $P_1, P_2, \ldots P_l$ constitute the group of proxy signcrypters. Each proxy signcrypter chooses secret key $x_{pj} \in Z_q^*$ and computes public key $y_{pj} = x_{pj}P$.

Let C be the unsigncrypter with secret key $x_c \in Z_q^*$ and public $y_c = x_c P$.

Let E and D be the encryption and decryption functions respectively and let $H_1 : \{0,1\}^* \rightarrow Z_q^*$, $H_2 : \{0,1\}^* \rightarrow G_1$ and $H_3 : G_2 \rightarrow \{0,1\}^n$ be the hash functions. System parameters are ($G_1, G_2, e, P, q, H_1, H_2, H_3, E, D$).

**Proxy Key Generation:** The original signcrypters make the signed warrant $m_W$ and broadcast to the proxy signcrypters. Through the following protocol, each proxy signcrypter $P_j$ gets a proxy key $S_{pj}$.

1. Each of the original signcrypters computes $S_{ai} = x_{ai} H_2(m_w)$ and broadcasts it all the proxy signcrypter.

2. The proxy signcrypter group verifies the correctness of each $S_{ai}$ by the equation

$$e(P, S_{ai}) = e(y_{ai}, H_2(m_w))$$

3. If each $S_{ai}$ is verified, then each proxy signcrypter $P_j$ computes $S_A = \sum_{i=1}^{n} S_{ai}$ and his proxy signcryption key $S_{pj} = S_A + x_{pj} H_2(m_W)$.

**Multi-Proxy Multi-Signcryption Generation:** One proxy signcrypter say $P_1$, may be designated as a clerk, whose task is to combine partial proxy signcryption to generate the final multi-proxy multi-signcryption.

1. Each $P_j$ selects an integer $t_j \in_R Z_q^*$, and computes $r_{pj} = e(P, y_c)^{t_j}$ and
2. $P_j$ broadcasts $r_{pj}$ to other proxy signcrypters, $P_1, P_2, \ldots, P_{j-1}, P_{j+1}, \ldots, P_l$.
3. Each $P_j$ computes $k = H_3(\prod_{j=1}^{l} r_{pj})$, $c = E_k(m)$, $r_p = H_1(c \| k)$

   and $u_{pj} = t_j P - r_p S_{pj}$, and sends $u_{pj}$ to the clerk $P_1$ as his partial proxy signcryption on m,
4. $P_1$ computes $u_p = \sum_{j=1}^{l} u_{pj}$, $S = lS_A$ and sends ($m_W$, S, c, $r_p$, $u_p$) to the b unsigncrypter C as multi-proxy multi-signcryption of m.

**Unsigncryption:** After ($m_W$, S, c, $r_p$, $u_p$), C computes

$$k = H_3(e(u_p, y_c) e(S, y_c)^{r_p} e(H_2(m_W), \sum_{j=1}^{l} y_{pj})^{r_p x_c})$$

and accepts the multi-proxy multi-signcryption if and only if $r_p = H_1(c \| k)$. She then recovers $m = D_k(c)$.

## 4 Our proposed scheme

To make the proposed scheme identity based, we introduce the Private Key Generator (PKG) along with original signcrypter group {$A_1, A_2, \ldots, A_n$}, proxy signcerypter group { $P_1, P_2, \ldots, P_l$ } and the unsigncrypter C.

**Set up:** Given a security parameter k, the PKG chooses group $G_1$ and $G_2$ of prime order q, a generator P of $G_1$, a bilinear map $e: G_1 \times G_1 \to G_2$ and hash functions $H_1: \{0,1\}^* \to G_1$, $H_2: G_2 \to \{0,1\}^n$ and $H_3: \{0,1\}^* \times G_2 \to Z_q^*$.
Then PKG chooses a secret key $s \in_R Z_q^*$ and computes $P_{Pub} = sP$. It also chooses a secure symmetric cryptosystem (E, D).
The system's public parameters are P = ($G_1, G_2$, n, e, P, $P_{Pub}$, $H_1, H_2, H_3$, E, D).

**Key generation:** Given an identity ID, the PKG computes user's public key $Q_{ID} = H_1(ID)$ and secret key $S_{ID} = sQ_{ID}$.
Each signcrypters $A_i$ has a public key $Q_{ID_{Ai}} = H_1(ID_{Ai})$ and secret key $S_{ID_{Ai}} = sQ_{ID_{Ai}}$.

Similarly each proxy signcrypter $P_j$ has a public key $Q_{ID_{pj}} = H_1(ID_{pj})$ and secret key $S_{ID_{pj}} = sQ_{ID_{pj}}$. Unsigncrypter; C has a public key $Q_{ID_c} = H_1(ID_c)$ and a secret key $S_{ID_c} = sQ_{ID_c}$.

**Proxy Keygenration:** To delegate the signcrypting capability to a group of proxy signcrypters, the original signcrypters make the signed warrant $m_W$. Which may include an explicit description of the delegation relation including the identity of the original signcrypters and the proxy signcrypters, the message to be signed, and so on. Then broadcasts $m_W$ to the l proxy signcrypters.

Each proxy signcrypter $P_j$ recover the proxy key $S_{pj}$ in the following manner:

1. Each $A_i$ computes $S_{Ai} = S_{ID_{Ai}} H_1(m_W)$ and broadcasts $S_{Ai}$ to the each proxy signcrypter.
2. The proxy signcrypter group verifies the correctness of each $S_{Ai}$ by the equation

    $$e(P, S_{Ai}) = e(P_{Pub}, Q_{ID_{Ai}} H_1(m_W)) \qquad i = 1, 2.......n$$

3. If all the above equation holds, the proxy signcrypter group computes $S_A = \sum_{i=1}^{n} S_{Ai}$.

    Each proxy signcrypter $P_j$ computes his proxy signcryption key $S_{pj}$ as
    $$S_{pj} = S_A + S_{ID_{pj}} H_1(m_W)$$

**Multi-Proxy Multi-Signcryption Generation:** Here we assume that one member of the proxy signcrypter say $P_1$, is the clerk who combines all partial proxy signcryptions.
.

1. Each $P_j$ selects $t_j \in_R Z_q^*$, and computes
    $$k_{1j} = e(P, P_{Pub})^{t_j} \in G_2 \text{ and } k_{2j} = e(P_{Pub}, Q_{ID_c})^{t_j} \in G_2$$

2. $P_j$ broadcasts $k_{1j}$ and $k_{2j}$ to the remaining proxy signcrypters.

3. Each $P_j$ now computes $k_1 = \prod_{j=1}^{l} k_{1j} \in G_2$ and

    $$k_2 = H_2(\prod_{j=1}^{l} k_{2j}) \in G_2, \quad c = E_{k_2}(m), \quad r_p = H_3(c, k_1), \quad u_{pj} = t_j P_{Pub} - r_p S_{pj} \in G_1,$$

    and sends $u_{pj}$ to the clerk $P_1$ as his partial proxy signcryption on m.

4. The clerk computes $u_p = \sum_{j=1}^{l} u_{pj}$, $S = lS_A$ and sends ($m_W$, S, c, $r_p$, $u_p$) to C multi-proxy multi-signcryption of m.

**Unsigncryption:** After receiving ($m_W$, S, c, $r_p$, $u_p$), C computes

$$k_1 = e(u_p, P)e(S,P)^{r_p} e(H_1(m_W), \sum_{j=1}^{l} Q_{ID_{pj}})^{r_p P_{Pub}} \text{ and}$$

$$k_2 = H_2(e(u_p, Q_{ID_c})e(S, Q_{ID_c})^{r_p} e(H_1(m_W), \sum_{j=1}^{l} Q_{ID_{pj}})^{r_p S_{IDc}})$$

accept the proxy signature if and only if $r_p = H_3(c, k_1)$.
Satisfied, she may recover the message as $m = D_{k_2}(c)$.

## 5 Security Analysis

**Verifiability:**

$$k_1 = \prod_{j=1}^{l} k_{1j} = \prod_{j=1}^{l} e(P, P_{Pub})^{t_j}$$

$$= e(\sum_{j=1}^{l}(u_{pj} + r_p S_{pj}), P)$$

$$= e(u_p, P)e(\sum_{j=1}^{l}(S_A + S_{ID_{pj}} H_1(m_W), P)^{r_p}$$

$$= e(u_p, P)e(S, P)^{r_p} e(H_1(m_W), \sum_{j=1}^{l} Q_{ID_{pj}})^{r_p P_{Pub}}$$

$$k_2 = H_2(\prod_{j=1}^{l} k_{2j}) = H_2(\prod_{j=1}^{l} e(P_{Pub}, Q_{ID_c})^{t_j})$$

$$= H_2(e(\sum_{j=1}^{l}(u_{pj} + r_p S_{pj}), Q_{ID_c}))$$

$$= H_2(e(u_p, Q_{ID_c})e(\sum_{j=1}^{l}(S_A + S_{ID_{pj}} H_1(m_W)), Q_{ID_c})^{r_p})$$

$$= H_2(e(u_p, Q_{ID_c})e(S, Q_{ID_c})^{r_p} e(H_1(m_w), \sum_{j=1}^{l} S_{ID_{pj}} Q_{ID_c})^{r_p}$$

$$= H_2(e(u_p, Q_{ID_c})e(S, Q_{ID_c})e(S, Q_{ID_c})^{r_p} e(H_1(m_W), \sum_{j=1}^{l} Q_{ID_{pj}})^{r_p S_{IDc}}$$

**Strong Unforgeability:** As for multi-proxy multi-signcryption, there are mainly four kinds of attackers: any third party, who do not participate the issue of the multi-proxy multi-signcryption; some proxy signcrypter, who plays an active part in signcryption process; the original signcrypter and the signcryption owner. Because the multi-proxy multi signcryption $u_p = \sum_{j=1}^{l} u_{pj}$ contains secret key information $S_{ID_{pj}}$ of each proxy signcrypter $P_j$ in the proxy multi-signcryption key generation phase, without secret key information $S_{ID_{pj}}$ of $P_j$, any third party, some proxy signcrypter, the signcryption owner and the original signcrypters cannot generate a valid multi-proxy multi-signcryption scheme by themselves.

**Strong Identifiably:** The unsigncrypter can distinguish proxy's normal signcryption from his multi-proxy multi-signcryption, because the multi-proxy multi-signcryption key is different from his own private key.

**Strong Nonrepudiation:** In our scheme, each proxy signcrypter $P_j$ cannot repudiate his participation on multi-proxy multi-signcryption while illegal attacker cannot claim that he is proxy signcrypter, because $u_p = \sum_{j=1}^{l} u_{pj}$ contains secret key information $S_{ID_{pj}}$ of each proxy signcrypter $P_j$, at the same time, warrant $m_W$ also contains identity information of $P_j$, in addition $k_2 = H_2(e(u_p, Q_{ID_c}) e(S, Q_{ID_c})^{r_p} e(H_1(m_W), \sum_{j=1}^{l} Q_{ID_{pj}})^{r_p S_{ID_c}})$ contains $Q_{ID_{pj}}$ and $m_W$.

**Confidentiality:** Because the secret key $k_2$ contains secret key information $S_{ID_c}$ of C, only C can compute $k_2$ and recover m.

**Prevention of Misuses:** No proxy signcrypter $P_j$ can repudiate his participation on multi-proxy multi-signcryption. In addition, $m_W$ includes the message to be signed, so our scheme can prevent the misuse.

**Public verifiability:** Any third party can verify the multisignature, so our scheme provides the public verifiability.

# 6  Conclusions

In this paper, based on Liu Jun Bao and Xiao Guo-Zhen scheme [9] we construct a new ID-based multi-proxy multi-signcryption scheme from pairing. As compared to Liu Jun Bao and Xiao Guo-Zhen scheme the proposed scheme provide the public verifiability of the proxy signature. Further the scheme is ID-based therefore; the key management problem is simplified.